\documentclass[conference]{IEEEtran}
\usepackage{graphicx}
\graphicspath{{./figures/}{./graphics/}{./graphics/logos/}}

\usepackage{graphicx}
\usepackage{algorithm}
\usepackage[noend]{algpseudocode}

\usepackage{breqn}

\usepackage{url}
\usepackage{hyperref}
\usepackage{braket}

\usepackage{subcaption}

\IEEEoverridecommandlockouts

\begin{document}
%
%
\title{Quantum Federated Learning: Analysis, Design and Implementation Challenges
}

\author{
Dev Gurung, Shiva Raj Pokhrel and Gang Li
}


\maketitle

\begin{abstract}

Quantum Federated Learning (QFL) has gained significant attention due to quantum computing and machine learning advancements. As the demand for QFL continues to surge, there is a pressing need to comprehend its intricacies in distributed environments. 
This paper aims to provide a comprehensive overview of the current state of QFL, addressing a crucial knowledge gap in the existing literature. We develop ideas for new QFL frameworks, explore diverse use cases of applications, and consider the critical factors influencing their design. The technical contributions and limitations of various QFL research projects are examined while presenting future research directions and open questions for further exploration.
\end{abstract}


\section{Introduction}\label{sec-intro}
Quantum computing is a transformative technology with the potential to revolutionize the digital world. 
It promises to reduce the computational complexity of machine learning tasks and improve model performance \cite{zhangFederatedLearningQuantum2022}. Large companies such as Google, IBM, and D-Wave have already introduced quantum computers, allowing machine learning based on quantum computing to be explored \cite{chenFederatedQuantumMachine2021a}.

QML (Quantum Machine Learning) has emerged as an exciting field that links quantum physics and machine learning and benefits from impressive advances in both domains. It has become an exciting area. Combining quantum computing and on-device machine learning has opened enormous opportunities for collaborative machine learning among devices, such as improving federated learning (FL)~\cite{pokhrelFederatedLearningBlockchain2020}. 


FL focuses on decentralized learning of global models while maintaining the privacy of local data while ensuring the confidentiality of local data. This privacy-preserving approach can be extended to the emerging field of QML to exploit the power of quantum computers.

QFL (Quantum federated learning) involves transforming classical federated learning (CFL) with relevant ideas from QML. In QFL, multiple devices with quantum computing capabilities work together to train a global model. 
Ideally, participating devices store their data in quantum states, called quantum data. However, in hybrid settings, classical data is encoded into quantum states for training and classification purposes in a hybrid state.
\begin{figure}[t]
\vspace{-6 mm}
    \centering
    \includegraphics[width=0.56\columnwidth]{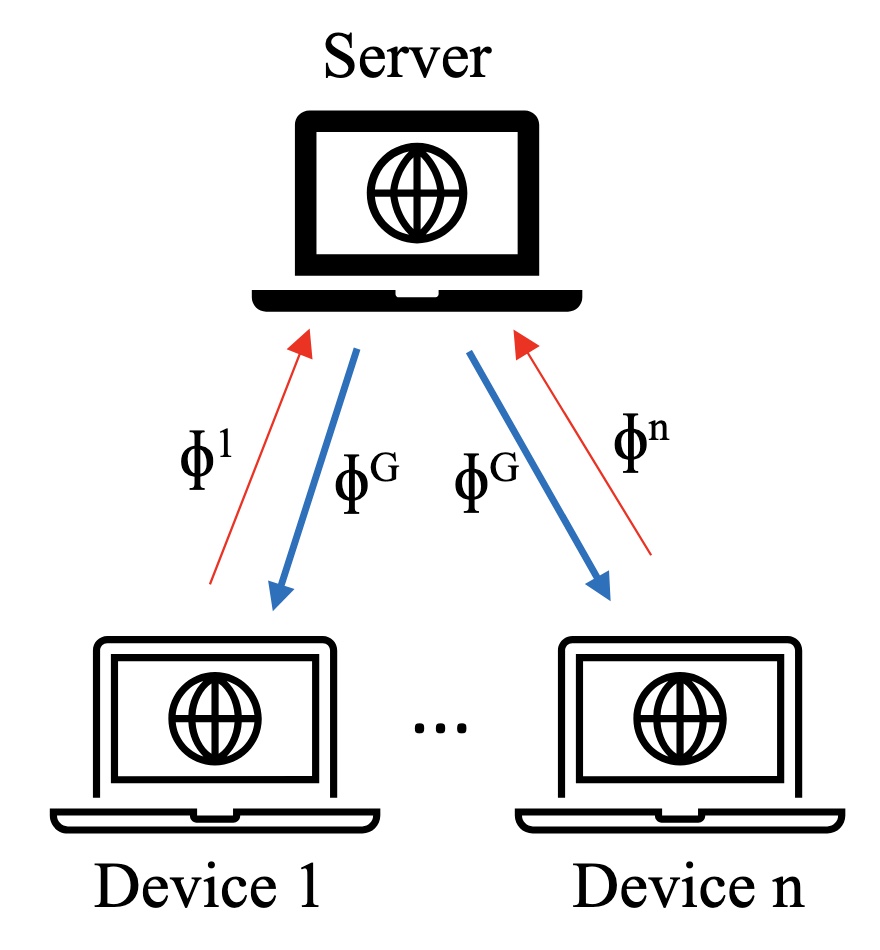}
    \caption{Understanding Quantum Federated Learning Architecture: The server broadcasts the global model $\phi^G$ to $n$ devices. Devices (1 \dots n) send back trained models $(\phi^1, ... , \phi^n)$ to the server.}
    \label{fig:qfl}
\end{figure}

\subsection{Motivation}
Advances in quantum technology have made QML an attractive option for federated learning, offering many benefits in privacy, security, and performance. This has made it a better option. However, some areas within QML still require further research and exploration. For example, quantum data availability and analyses are still in their infancy. In addition, the current availability of quantum devices is limited to noisy mid-scale quantum (NISQ) devices. Classical computers have certain computational limitations. These limitations of machines also limit machine learning capabilities. 

Although Moore's law has traditionally predicted exponential growth in computing capacity, the development of classical computing is now stalling. This limitation highlights the need to explore alternative approaches like QML to overcome computational bottlenecks and unlock new possibilities to solve complex problems. 
Therefore, addressing hardware challenges and moving towards QML can open up new opportunities to overcome limitations and advance machine learning.

Our primary focus in this work is to fill the gap of extremely limited QFL studies and their designs. The main objective of this work is to develop a holistic understanding of the landscape of QFL by evaluating existing methodologies, identifying challenges and opportunities, and ultimately guiding novel research directions.

\subsection{Contribution}
The main contributions of this work are as follows:

\begin{enumerate}
\item We provide a comprehensive overview of the current state of research in QFL 
followed by discussing existing approaches and limitations, and identifying future research directions.

\item We implement two proof-of-concept simulations considering different factors such as encoding methods, number of qubits in the quantum circuit, number of layers, etc., and their impact on overall system performance.




\end{enumerate}

\section{Background}
Quantum computing is a field of study that uses principles from quantum mechanics to perform computational tasks. It takes advantage of operations such as the quantum Fourier transform and amplitude amplification, potentially providing significant computational advantages over classical systems \cite{panExperimentalQuantumEndtoend2023}.

At the core of quantum computing is the concept of a qubit, which serves as the basic unit of information storage. Unlike classical bits, which can only represent values of 0 or 1, qubits can exist in a superposition of states, allowing for a broader range of importance due to quantum mechanical phenomena.
A \textit{qubit} is the fundamental unit of information, and unlike bits, which can only have a value of 0 or 1, 
qubits exhibit unique properties of quantum mechanics, such as \textit{superposition} and \textit{entanglement},
that make quantum computing more powerful than classical computing.
A qubit has two basis states, denoted as $\ket{0}$ and $\ket{1}$. 
However, due to superposition, a qubit can exist in a linear combination of these basis states, represented as:
$$\ket{\psi} = \alpha \ket{0} + \beta \ket{1}$$
Here, $\alpha$ and $\beta$ are amplitudes associated with the basis states and satisfy condition $|\alpha|^2 + |\beta|^2 = 1$.

The number of qubits required in a quantum computer depends on the computational problems being addressed.
The transmission medium for quantum information, represented by qubits, is called a quantum channel. 
It plays a crucial role in quantum communication and
enables the transfer of quantum states between different components or systems.

The current generation of quantum computers have
a limited number of qubits and are susceptible to errors are termed
Noisy Intermediate-Scale Quantum (NISQ) 
 computers.

 Although not fault-tolerant, NISQ computers offer opportunities to explore quantum algorithms and applications. 
 However, large-scale quantum algorithms are still expected to be executed on fully fault-tolerant quantum computers, which are still under development.

\textit{Quantum Neural Network (QNN)} is a counterpart to classical neural networks that incorporates the
fundamental principles of learning from training data. 
Similarly to classical neural networks, a QNN takes input $x$ and computes an output $y$. However, instead of processing classical data, 
QNNs operate on quantum data, which are represented in qubits. 
QNNs leverage the principles of quantum mechanics to evaluate and compute the output based on the quantum input. Quantum perceptron is the fundamental building block of QNN.
It is analogous to the classical perceptron in traditional neural networks but operates using quantum principles.

When observed or measured, a qubit collapses to one of the basis states with corresponding probabilities determined by the amplitudes. 
These probabilities can be utilized for statistical estimation after multiple repeated observations.
The significant advantage of qubits over classical bits is their ability to represent an exponentially large amount of information. 
This is achieved through the phenomenon of \textit{entanglement}, 
where qubits correlate. With $n$ entangled qubits, the system can exhibit $2^n$ basis states, 
enabling powerful computational capabilities and leveraging the superposition effect.

Computation on qubits is achieved through the utilization of quantum logic gates. 
Some common examples of quantum logic gates include Pauli-X, Pauli-Y, Pauli-Z, Hadamard, and Controlled-Not gates.
An essential characteristic of quantum logic gates is their reversibility,
which sets them apart from classical gates such as AND and OR. 
In classical computing, these logic gates can lead to information loss, whereas in quantum computing, quantum gates are designed to preserve information and enable reversible computations.

To explore and exploit QNNs, we require quantum data, which can be obtained by either encoding classical data into quantum states or directly generating quantum data from quantum computers. Currently, the most common approach is to encode or transform classical bits into their qubit representation.
For this purpose, a $n$ -qubit state $\ket{\psi}_n$ can be employed to represent a superposition of $2^n$ basis states in the Hilbert space $\mathcal{H}^{2^n}$. Mathematically, it can be expressed as:
$$\ket{\psi}_d = \sum_{i=1}^{2^n} \beta_i \ket{S_i}, $$ 
where, $\beta_i$ represents complex amplitudes and $\ket{S_i}$ denotes the basis state in the Hilbert space $\mathcal{H}^{2^n}$. Using quantum data, QNNs can effectively operate on these quantum states to perform computations and make predictions in QML tasks.

\subsection{Data Encoding}
Data encoding is the process of transforming classical information into quantum states that a quantum computer can manipulate. 
Different encoding methods are applied to classical data to convert them into quantum data.
Two common approaches to encoding classical data into quantum states are amplitude encoding and binary encoding. 
Amplitude encoding involves representing data through the amplitudes of quantum states, while binary encoding assigns information or bits to the states of qubits.
In QFL, selecting the most
suitable encoding approach is crucial for accurately representing classical data as quantum states. 
The choice of encoding method can impact the performance and efficiency of quantum algorithms.
Binary encoding is a suitable choice when data is processed
using arithmetic computations in quantum algorithms. 
In binary encoding, information or bits are represented by the
states of qubits, allowing for straightforward manipulation and operations on the encoded data.
Analog encoding is a preferred method when data are mapped into the Hilbert space of a quantum device. 
In this encoding scheme, the information is represented by the amplitudes or continuous values of the quantum states. 

In our sample designs, we implement three simple types of data encodings: $vanilla, mean$, and $half$.
With $vanilla$, data remain mostly unchanged, while with the $mean$ approach, data are centered with mean subtraction. For $half$ encoding, the data are subtracted by half.

\subsection{Variational Quantum Circuits (VQC)}
Variational Quantum Circuits (VQCs) are quantum circuits that possess adjustable parameters \cite{chenFederatedQuantumMachine2021a}. In the general structure of a VQC, there is an encoder responsible for transforming classical data into quantum states. The quantum gates are then applied to these states, allowing quantum operations and computations. Additionally, VQCs feature a learnable unit block where the parameters can be trained, similar to the weights in classical neural networks. Finally, the information processed within the quantum circuit is extracted and converted to a classical format for further analysis or utilization.

VQCs provide a flexible framework for performing quantum computations on encoded data and enable the optimization of the circuit's parameters to enhance performance in QML tasks. Also known as
parametrized quantum circuits, they form the foundation of near-term QML algorithms. 
These circuits employ gates with adjustable parameters that can be tuned during the learning process to optimize the performance of quantum models. 
Parameterized quantum circuits enable the exploration and optimization of QML models.

\subsection{Quantum Federated Learning (QFL)}
In QML, the objective is to find a quantum channel, denoted $M$, that takes an input
$x$ and predicts the corresponding class. This is achieved by tuning the variational parameters, denoted as $w$, of a quantum
channel, specifically $M_w$, to minimize the average loss function
$\mathcal{L}$ over a given data set $D$ containing $n$ samples
\cite{zhaoExactDecompositionQuantum2022a}:

\begin{equation}
\min_{w} \frac{1}{n} \sum_{(x, y) \in D} \mathcal{L}(w, x, y)
\end{equation}

To solve this optimization problem, an iterative approach is employed using gradient descent with a learning rate $\eta$ is employed. At a specific time $t$, the variational parameters are updated as follows:

\begin{equation}
w_t = w_t - \eta \Delta_w \frac{1}{n} \sum_{(x, y) \in D} \mathcal{L}(w,x,y)
\end{equation}

Here, $x$ represents the quantum state which means the input data. The goal is to iteratively adjust the variational parameters to minimize the average loss and enhance the performance of the quantum channel for prediction tasks in QML.

\section{Literature Review}

\begin{table*}[!h]
\centering
\begin{tabular}{p{2cm}|p{2cm}|p{6cm}|p{6cm}}
\hline

\textbf{References (First-Author)} & \textbf{QFL Context} & \textbf{QFL Design Idea} & \textbf{Remarks} \\
\hline
\hline

Xia  (2021) \cite{xiaQuantumFedFederatedLearning2021a} & QuantumFed & Proposed collaborative QFL framework. & Limited scalability to large-scale quantum systems. \\
\hline
Chehimi  (2022) \cite{chehimiQuantumFederatedLearning2022} & QFL Framework & Comprehensive QFL framework for quantum data. & Lack of experimental validation on real quantum hardware. \\
\hline
Huang  (2022) & QFL algorithm & Efficient QFL algorithm for VQA. & Privacy issues in the use of Quantum Data \\
\hline
Li  (2021) \cite{liQuantumFederatedLearning2021} & Private QFL & Private framework for blind quantum computing. & Reliance on the trusted quantum server for computation. \\
\hline
Xia  (2021) \cite{xiaDefendingByzantineAttacks2021} & Comparative study & Comparative study: Byzantine attacks over classical FL vs. QFL. & Limited analysis of Byzantine attacks in QFL; Only use of Gaussian distribution. \\
\hline
Zhao  (2022) \cite{zhaoExactDecompositionQuantum2022a} & Non-IID data & Addressed non-IID data performance in QFL. & Additional inference step compared to regular FL averaging. \\
\hline
Zhang  (2022) \cite{zhangFederatedLearningQuantum2022} & Quantum Secure aggregation  & Secure aggregation scheme in QFL using entangled qubits. & Limitation to practical implementation. \\
\hline
Chen  (2021) \cite{chenFederatedQuantumMachine2021a} & Hybrid QFL & Hybrid quantum-classical federated training. & Challenges in optimizing the integration of quantum and classical models. \\
\hline
Yamany  (2021) \cite{yamanyOQFLOptimizedQuantumBased2021} & Optimized QFL & Optimized QFL for automatic adjustment of hyperparameters. & Limited evaluation on the robustness against advanced adversarial attacks: inversion attacks. \\
\hline
Yun  (2022) \cite{yamanyOQFLOptimizedQuantumBased2021} & Slimmable QFL & Slimmable QFL that addresses communication and energy limitations. & Require further investigation under communication channel conditions and non-IID global data.\\

\hline
Gurung (2023) \cite{gurungSECURECOMMUNICATIONMODEL2023a} & PQC-QFL Model  & Post-quantum secure communication for QFL with Dynamic Server selection & Experimental analysis and design \\
\hline
Ren  (2023) \cite{renQuantumFederatedLearning2023} & QFL survey  & Thorough holistic examination of QFL. & Lacking experimental analysis. \\\hline
\end{tabular}
\caption{Literature on QFL analysis, design, and implementation}
\label{table:literature_review}
\end{table*}

\subsection{QFL Frameworks}
Xia \textit{et al.} \cite{xiaQuantumFedFederatedLearning2021a} presented a federated learning framework for collaborative quantum training. 
In this framework, multiple quantum nodes with local data collectively train a global quantum neural network model.
Data plays a critical role in federated learning and is equally essential in QFL. 
Without quantum data, QFL would not be possible.
Chehimi \textit{et al.} \cite{chehimiQuantumFederatedLearning2022} proposed a comprehensive quantum federated framework that operates exclusively on quantum data. 
They claim superior performance compared to a centralized pure QML setup.
Huang \textit{et al.} \cite{huangQuantumFederatedLearning2022} introduced a QFL algorithm to improve the communication efficiency of Variational Quantum Algorithms (VQA) using decentralized non-IID quantum data.
In blind quantum computing, Li \textit{et al.} \cite{liQuantumFederatedLearning2021} presented a private distributed learning framework. 
The proposed method allows each client to take advantage of the computational power of a quantum server. 
Differential privacy techniques are employed to ensure privacy, eliminating the need for trust in the server.
However, a challenge with this approach arises when clients cannot prepare single qubits for transmission to the server.
This limitation may arise in scenarios involving classical clients or other circumstances where the capability to prepare
qubits is not available.
In terms of security, Gurung et al. \cite{gurungSECURECOMMUNICATIONMODEL2023a} proposed the postquantum cryptography (PQC) QFL model to protect communication between clients and server from quantum threats. The paper also proposed dynamic server selection to improve system efficiency.
As the work presented was as a proof of concept, further extensive experimental analysis is required.

\subsection{Types of Applications}

Zhao et al. \cite{zhaoExactDecompositionQuantum2022a} addressed performance issues in QFL caused by identically distributed non-independent (non-IID) data. The proposed solution involves decomposing the global quantum channel into locally trained channels for each client using local density estimators. Their framework called \textit{qFedInf}~\cite{zhaoExactDecompositionQuantum2022a} achieves a single-shot communication complexity, eliminating the need for multiple communication rounds as in regular federated learning. However, it involves two phases: the training phase and the inference phase. The inference phase requires sending the trained channels and density estimators to the server, introducing an additional step compared to traditional FL averaging. The practicality of \textit{qFedInf}~\cite{zhaoExactDecompositionQuantum2022a} warrants further investigation.

Zhang \textit{et al.} \cite{zhangFederatedLearningQuantum2022} proposed a Quantum Secure Aggregation (QSA) scheme for secure and efficient aggregation of local model parameters in federated learning using quantum bits. Leveraging entangled qubits improves the computational complexity of transmitting and aggregating model parameters. The study considers both centralized and decentralized architectures in a horizontal federated learning setting. The authors claim that their QSA scheme achieves "perfect secrecy" compared to classical secure aggregation methods, such as differential privacy and homomorphic encryption, while also offering better performance. However, the applicability of the proposed scheme is based on the feasibility of teleporting entangled states and implementing quantum gates over longer distances.

Chen \textit{et al.} \cite{chenFederatedQuantumMachine2021a} presented federated training based on hybrid quantum-classical machine learning models. Their approach combines a quantum neural network (QNN) with a trained classical convolutional model, employing a hybrid quantum-classical transfer learning architecture. This work highlights the need to explore different types of QML approaches.

In the context of intelligent transportation systems, Yamany \textit{et al.} \cite{yamanyOQFLOptimizedQuantumBased2021} proposed an optimized quantum-based federated system to defend against adversarial attacks. The approach involves adjusting hyperparameters for federated learning using various adversarial attacks tailored to autonomous vehicles. The tuning of hyperparameters, including the federated learning model, is accomplished using the Quantum Behaved Particle Swarm Optimization (QPSO) algorithm.
Yun \textit{et al.} \cite{yunSlimmableQuantumFederated} presented slimmable QFL, which is a dynamic QFL framework that addresses two specific challenges: time-varying communication channels and limitations in computing energy. 

\section{QFL Operationalization Challenges}
One of the important aspects in the context of QML is the development of VQAs.
VQAs aim to leverage the capabilities of current NISQ devices to achieve a quantum advantage in solving computational problems. 
However, NISQ devices suffer from limitations, such as high error rates or "noise" in their computations, which hinder their feasibility for real-life applications.
With poorly performing VQAs, efficient QFL algorithms is hard to achieve.

\subsubsection{Enhanced Data Encoding for QFL}
Data encoding plays an important role in current QML approaches, as classical data need to be converted to quantum data. 
With improved data encoding approaches, the performance of the entire QFL framework could be enhanced. 
However, there is still work to go in this direction to standardize or advance the technique of data encoding.

\subsubsection{QFL Driven Parameterized Circuits Tuning}
When choosing a parameterized quantum circuit for machine learning applications, it is important to consider its generalizability. 
The circuit should have the ability to generate diverse subsets of states within the Hilbert space to capture a wide range of quantum patterns. 
Additionally, the circuit should be capable of entangling qubits, as entanglement plays a crucial role in quantum computation and can enhance the representation power of the circuit. By selecting parameterized circuits that meet these criteria, we can ensure that the chosen circuit is well-suited for QFL tasks.

\subsubsection{Addressing Hardware Limitations for QFL}
There are several challenges related to hardware in the context of QML. 
One challenge is the presence of noise in quantum systems, which can affect the accuracy and reliability of computations. 
Additionally, there is concern about possible malicious updates from devices participating in QML, which can compromise the security and integrity of the learning process.
Another aspect to consider is the quantum nature of communication and computation, which introduces unique considerations compared to classical approaches. 
Quantum systems operate according to the principles of superposition and entanglement, which can enable powerful computations, but also require specialized hardware and protocols.
To address these hardware challenges, researchers are exploring various solutions for QFL--extending Quantum Stochastic Gradient Descent (QSGD).

\subsubsection{Decentralization of QFL}

By distributing the training process and reducing the dependence on a central server, decentralized FL can provide advantages such as improved privacy, reduced communication overhead, and improved scalability.
To achieve this, it is crucial to explore the integration of technologies like blockchain with QFL networks. 
The decentralized and immutable nature of blockchain can provide a transparent and secure framework for managing QFL processes,
ensuring data integrity, and enabling trusting collaborations between participants.

\subsubsection{Security and Privacy of QFL}
Security and privacy are critical considerations
in the context of QFL systems.
One aspect is privacy-preserving techniques, which aim to protect sensitive information during the federated learning process. 
These techniques ensure that participant data remain confidential and secure, while still enabling collaborative learning.
Another significant concern is the vulnerability to Byzantine attacks in QFL environments. 

\subsubsection{Understanding QFL Performance}
The performance of QFL systems encompasses various aspects that impact their effectiveness and efficiency.
Computational cost is a significant consideration, particularly in terms of communication and training time.

Communication in QFL involves transmitting large volumes of data, such as model weights or gradients, which can introduce overhead. 
Furthermore, training deep QNNs can be computationally expensive, and as the advancement of classical computing slows down, alternative approaches such as QFL become more compelling.
However, it should be noted that the cost functions commonly used in classical neural networks may not be easily computable with quantum operations, which poses challenges in developing appropriate quantum cost functions.
Training efficiency and cost are also crucial factors. 

CFL algorithms may face increasing computational costs and efficiency challenges as the size of the dataset grows, while quantum computers can offer improved training efficiency.
However, it is important to consider the specific challenges in QFL, such as finding the appropriate quantum generalization of neural network architectures, 
optimization algorithms, and loss functions for quantum data.
Furthermore, training on multiparty quantum devices is an essential challenge in order to take advantage of QFL and improve training efficiency.
It is worth noting that QNNs accelerate training over classical neural networks 
highlighting one of the main advantages of employing QFL among others.

\section{Future Directions}
\subsubsection{Emerging Techniques}
Causal Inference, which aims to understand cause-and-effect relationships from observational data can be incorporated into QFL, 
 to derive new insights and make informed decisions based on the causal relationships discovered within the distributed quantum data.
Another valuable integration is Knowledge Distillation, a technique that enables the transfer of knowledge from a complex model (teacher) to a simpler model (student). 
By leveraging knowledge distillation in QFL, 
quantum models can benefit from the expertise and insights gained by larger, 
more complex models, leading to improved performance and generalization.
Furthermore, the architecture of blockchain-based QFL systems remains an open area for exploration. 
Extensive research is needed to understand the design considerations, consensus mechanisms, and data management approaches that can optimize
the integration of blockchain technology with QFL. 
This would ensure the secure and efficient coordination of quantum
computations and data sharing among participants in a decentralized
manner.
By integrating these new techniques into the QFL framework and addressing the architectural aspects of blockchain-based QFL systems, researchers can unlock new possibilities and advances in QML, ultimately leading to more robust and effective quantum learning systems.

\subsubsection{QFL gains over CFL}
Quantum computing offers the potential for significant
advantages over classical machine learning algorithms. 
Quantum algorithms can exploit quantum superposition
and entanglement to perform computations in parallel, 
leading to potentially faster processing and
improved efficiency. 
QML
algorithms have the ability to handle large-scale data and complex patterns more effectively, enabling enhanced learning capabilities. 
Additionally, quantum algorithms can leverage quantum state transformations and quantum interference to extract information and make more accurate predictions. 
Although the quantum advantage in QFL is still an ongoing area of research and development, 
the unique properties of quantum computing hold promise for tackling challenging computational tasks and unlocking
new opportunities in data analysis and pattern recognition.

\subsubsection{Gate-based vs Pulse Based QFL}
The choice between gate-based and pulse-based quantum computing
approaches is an important consideration in quantum information processing. 
Gate-based quantum computing relies on a series of discrete gates, 
such as single-qubit gates and two-qubit gates,
to manipulate quantum states and perform computations. 
This approach offers precise control over quantum operations and allows the implementation of various quantum algorithms. 
On the other hand, pulse-based quantum computing utilizes
continuous-time control over individual qubits using microwave pulses.
This approach offers more flexibility and scalability in implementing
quantum operations but requires careful calibration and mitigation of errors arising from noise and imperfections in the hardware. 
The decision between gate-based and pulse-based approaches depends on
the specific requirements of the quantum task at hand, 
the available hardware technology, and the trade-offs between control
precision and scalability.
Ongoing research and experimental studies, such as the work by Pan et al. \cite{panExperimentalQuantumEndtoend2023}, contribute to the advancement of our understanding of the capabilities and limitations of the gate-based and pulse-based quantum computing
paradigms, guiding the development of future quantum technologies.

\subsubsection{QFL Implementations}
The implementation of QFL algorithms and applications faces various
challenges, particularly in the NISQ era. 
NISQ devices, which are the current generation of quantum computers, 
have limitations in terms of the number of qubits and computing power.
These devices are error-prone and not fault-tolerant, which restricts their application to small-scale computations. 
The practical implementation of quantum key distribution (QKD) is also
an area of focus. 
The availability of fault-tolerant quantum computers
is a significant milestone in achieving quantum advantage,
which refers to the speed-up of applications compared to
classical counterparts. 
Training, accuracy, and efficiency of VQAs are important considerations for harnessing the capabilities of NISQ devices. 
Ongoing research, such as the work by Cerezo et al. 
\cite{cerezoVariationalQuantumAlgorithms2021}, 
aims to address these implementation challenges and explore the 
the potential of near-term quantum computers in practical applications.
\subsubsection{Quantum Data for QFL}
The study of quantum data is an emerging field that explores the unique
characteristics and properties of data in the quantum realm. 
Quantum data for QFL refers to information encoded in distributed quantum systems, 
such as qubits, that can exist in superposition states and exhibit entanglement. 
Understanding quantum data for QFL is crucial for developing underlying algorithms
and applications that can leverage these quantum properties to solve
complex problems more efficiently. 
The field of quantum data encompasses topics such as quantum data
representation, quantum data storage and retrieval, quantum data
compression, and quantum data processing. 
Actively investigating techniques and methodologies
to extract meaningful insights from quantum data and harness its
potential for various domains, including QML and QFL has been impeding progress. 
\subsubsection{QFL Aggregation}
In terms of global optimization approaches in quantum federated learning, 
several techniques and methodologies are being explored. 
One such approach is asynchronous distributed learning, 
where local training processes occur independently with intermittent communication and synchronization. 
Such an approach under QFL will enable for faster convergence and better scalability in
large-scale federated systems. 
Another important aspect is privacy-preserving approaches, where
techniques like differential privacy and blind quantum computation are
employed to protect the privacy of data during the federated learning process. 
Integrating these privacy-preserving methods into quantum federated
learning can enhance the security and privacy of QFL models.
Additionally, advanced aggregation schemes are being developed in QFL to
address the compromise of global models that may occur due to the
upload of corrupt trained models from the clients. 
Such QFL aggregations would improve the robustness and accuracy of
the global model by effectively handling and filtering of
malicious unreliable updates.

\section{Proof of Concepts}
We develop two \textbf{proofs-of-concept (PoC)} to demonstrate the feasibility of what we have discussed so far and show the evaluations of QFL with quantum implementation.

\subsection{Settings of \textbf{PoC 1} and \textbf{PoC 2}}
In \textbf{PoC 1}, the variational classifier is implemented using pennylane library, a cross-platform python library for programming quantum computers.
With the pennylane quantum computation, quantum circuits are modeled as quantum computing nodes, implemented as special Python functions or quantum functions that operate with a number of gates.
To simulate the federated settings, for $it$ number of iterations, $n$ number of devices are trained collaboratively. 
To observe the performance of the devices collectively, their performance in accuracy is averaged, and communication performance is observed.

Different from \textbf{PoC 1}, in \textbf{PoC 2}, the Tensor circuits over the JAX framework are developed. Along the lines of implementing \textbf{POC 1} in pennlylane, a VQC is created for the quantum circuit and applied with a series of gates that takes the input quantum circuit $c$ with parameters $params$ and the number of layers $k$.
The first MNIST data set is shared amongst the $n$ number of devices with only $3$ classes for each device.
Different encoding types, such as \textit{vanilla, mean}, and \textit{half}, are applied to encode classical data to quantum states.
The images are also resized to fit the image size into the quantum circuit. 
Then, each device performs local training and updates the local model.
Whereas the server applies averaging of parameters from all devices, and each device's local parameters are updated with these
new global parameters, and training continues for a certain number of communication rounds.
Finally, the output result of the quantum circuit is read and converted into its classical representation.

\subsection{\textbf{PoC 1.} QFL experiments with Pennylane}
\begin{figure}[t]
    \centering
    \begin{subfigure}[t]{0.47\columnwidth}
    \centering
    \includegraphics[width=\columnwidth]{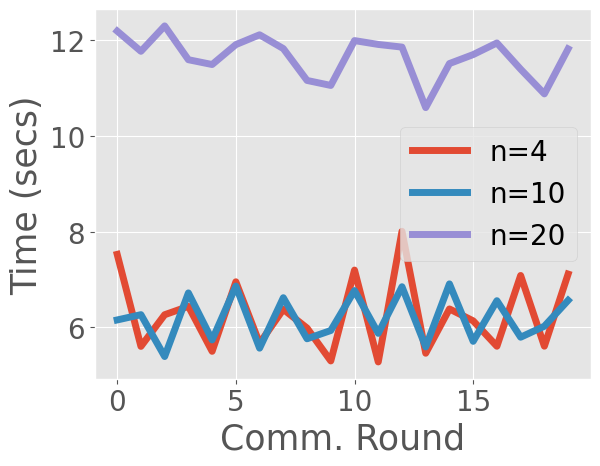}
    \label{fig:comm_time_noofdevices_pennylane}
    \end{subfigure}
    \begin{subfigure}[t]{0.475\columnwidth}
    \centering
    \includegraphics[width=\columnwidth]{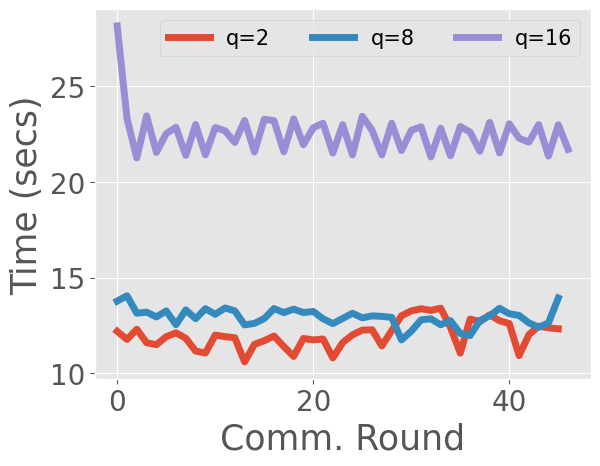}
    \label{fig:comm_time_qubits_pennylane}
    \end{subfigure}
    \caption{Performance with number of devices and qubits}
    \label{fig:comm_time_performance}
\end{figure}

For this PoC, we develop a module on the top of the penny-lane library using IRIS dataset. 
The variable classifier for the IRIS data set\footnote{https://github.com/PennyLaneAI/qml} is extended to simulate federated setting, observing the behavior of a quantum variational classifier in the QFL with the scaled IRIS dataset, ``default.qubit" device is employed from pennylane. 
From figure \ref{fig:comm_time_performance}, we can observe impact on overall communication delay of QFL based on number of devices $n$ and number of qubits $q$. 
Its quite clear that more number of devices imposes more delay to the overall system performance. 
This emphasizes the need to model QFL systems in such a way that the number of devices selected should be optimal.
A similar pattern can be observed with the number of qubits used impacting the performance differently, with more qubits imposing more delay to the system and vice versa.
Thus, to gain optimal performance of the QFL framework, these factors need to be especially explored and studied to
pinpoint critical reasons for their impact on the system.

\subsection{\textbf{PoC 2.} QFL Experimentation with Jax}
For the QFL experimentation, the implementation\footnote{https://github.com/s222416822/PQC-QFL-Model} is slightly modified to understand the nature and impact of different factors on QFL.
Three different types of data encoding tested are $vanilla, mean$ and $half$.

\begin{figure}[!h]
    \centering
    \includegraphics[width=0.69\columnwidth]{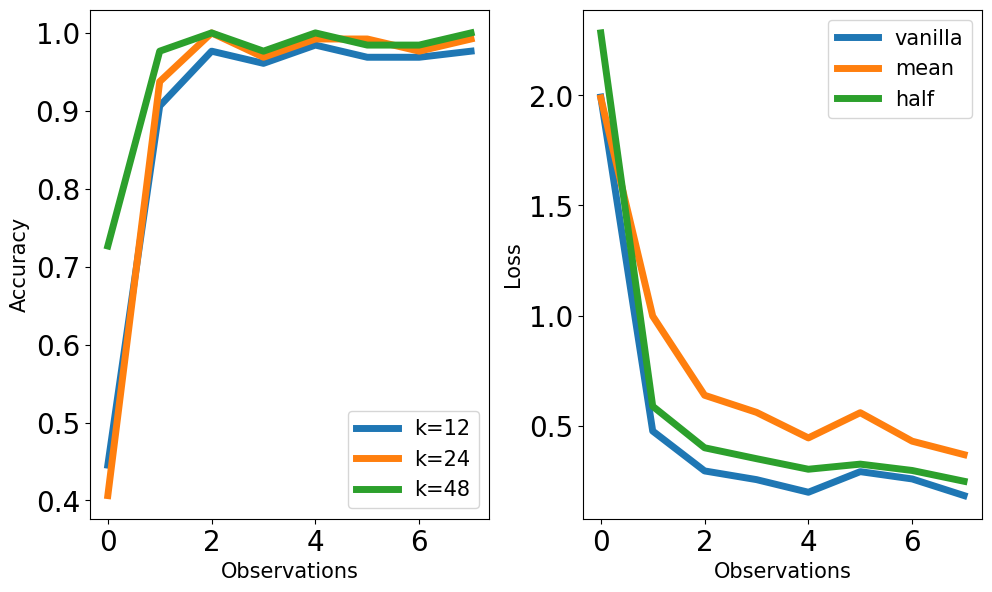}
    \caption{Training Performance}
    \label{fig:train_performance}
\end{figure}

Figure \ref{fig:train_performance} shows the impact of the number of layers $k$ and the types of encoding on the QFL training performance.
We can see in Figure~\ref{fig:train_performance} that the higher the $k$, the better the performance, while, among others, the $mean$ encoding performs the best and $vanilla$ shows the best performance in our QFL setup.
The reason behind this could be that the $mean$ and $half$ encoding suffers from a better representation of the data of classical data in quantum states. 
This implies why data encoding mechanism is such an important aspect to modeling QFL algorithms, which requires further research in the area.

   

\begin{figure}[t]
    \centering
    \includegraphics[width=0.69\columnwidth]{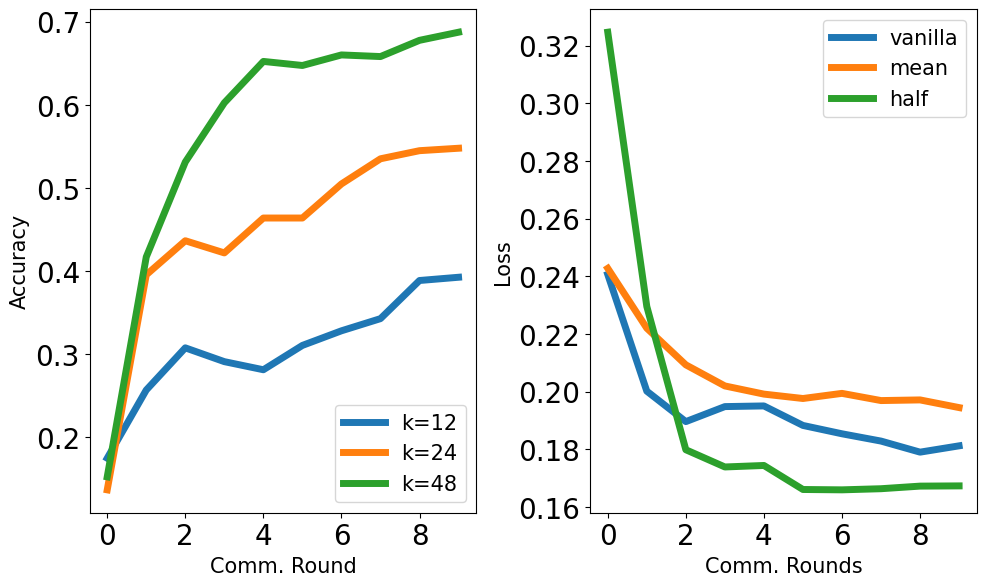}
    \caption{Test Performance}
    \label{fig:test_k}
\end{figure}

Figure \ref{fig:test_k} shows that test performance is higher with $k = 48$ and less with $k = 12$, while the half-encoding observes the lowest loss with a higher communication round. 
This is due to the increased complexity and computation with more layers in the quantum circuit.
In terms of the impact of the data encoding approach on QFL, the $half$ encoding approach seems to perform better than others, while the $mean$ approach is lagging in performance.

\section{Conclusion}
We have explored the current state-of-the-art in these fields, identified key challenges and limitations, and discussed potential future directions. 
Advances in quantum computing and machine learning are promising for solving complex problems and improving computing capacity. 
QFL, in particular, offers a distributed approach that combines the power of quantum systems with privacy and security considerations. 




\end{document}